# Non static local string in higher dimensional gravity

[1]F. Rahaman, [2]B.C.Bhui and [3]M. Kalam

[1] Department of Mathematics, Jadavpur University,
Kolkata-700032, India
[2] Dept. of Maths., Meghnad Saha Institute of Technology,
Kolkata-700150, India
[3] Dept. of Phys., N.N.College for Women, Kolkata-700064, India

[1]E-mail: farook_rahaman@yahoo.com

**Abstract:** We analyze the space-time structure of local gauge string with a phenomenological energy momentum tensor, as prescribed by Vilenkin in an arbitrary number of space-time dimensions with a non-zero cosmological constant Λ. A set of solutions of full non-linear Einstein's equations for the interior region of such a string are presented.

Finding a theory that unifies gravity with other forces in nature an elusive goal for theoretical physicists. At first, Kaluza-Klein ( K-K) have shown how gravity and electro magnetism can be Unified from Einstein's field equations generalized to five dimensions [1]. After that, higher dimensional cosmological models have been quite often present in scientific research. In recent years, there has been renewed interest in 'brane-world' models in which the universe is represented by a ( 3 + 1) dimensional subspace ( 3-brane) embedded in a higher dimensional (bulk) space-time [2]. The idea of brane world may resolved the challenging problem in theoretical physics namely the unification of all forces and particles in nature. According to Randall and Sundrum [3], it is possible to have a single massless bound state confined to a domain wall or 3-brane in five dimensional non factorizable geometries. The brane is pictured on a domain wall propagating in a 5-dimensional bulk space-time. They have shown that this bound state corresponds to the zero mode of Kaluza-Klein dimensional reduction and is related to the four dimensional gravity [3]. Recently Cohen and Kaplan [4] have shown that global string has a curvature singularity at a finite distance from the string core. They argued that the singularity can provide an effective compactification of extra dimension. Gregory [5] has shown that a non-singular global string exists in presence of negative cosmological constant. In her solution, the extra dimension are infinite and strongly warped as in Randall and Sundrum model.
Olasagasti and Vilenkin [6] obtained solutions of Einstein equations for global defects in higher dimensional space-time with a non zero cosmological constant.

In this report, we shall explore solutions of Einstein equations for local string in a higher dimensional space-time with a non-zero cosmological constant.

We assume an infinite long straight string, characterized by the energy momentum tensor components $T_t^t = T_{zi}^{zi} \neq 0$ and all other $T_i^k$ 's are zero [7]. Because of relevance of the brane scenerio, in the context of cosmic string, it becomes necessary to investigate whether a local gauge string can give rise to essential solutions of gravitational field equations in a higher dimensional space-time with a non zero cosmological constant. We shall adopt the following ansatz for the metric

$$ds^2 = e^{2A(r)} [dt^2 - e^{2b(t)} dz_i^{\,2}] - dr^2 - C^2(r) d\theta^2 \qquad \ldots(1)$$

where $i = 1, 2, \ldots p$ are the spatial coordinates in our lower dimensional space-time [5]. The local string is characterized by an energy density and stresses along the symmetry axes given by

$$T_t^t = T_{zi}^{zi} = F(r,t) \qquad \ldots(2)$$

and 
$$T_r^r = T_\theta^\theta = 0 \qquad \ldots(3)$$

The system of equations for local string are

$$(C^{11}/C) + p[A^{11} + (A^1)^2 + (A^1 C^1 / C)] + \tfrac{1}{2} p(p-1)[(A^1)^2 - (\dot{b})^2 e^{-2A}]$$
$$= 8\pi G\, F(r,t)\, e^{2A} - \Lambda \qquad \ldots(4)$$

$$p[\ddot{b} + (\dot{b})^2 e^{-2A}] - p(A^1)^2 - (p+1)(A^1 C^1 / C) - \tfrac{1}{2} p(p-1)[(A^1)^2 - (\dot{b})^2 e^{-2A}]$$
$$= -\Lambda \qquad \ldots(5)$$

$$p[\ddot{b} + (\dot{b})^2] e^{-2A} - (2p+1)(A^1)^2 - (p+1) A^{11} - \tfrac{1}{2} p(p-1)[(A^1)^2 - (\dot{b})^2 e^{-2A}]$$
$$= -\Lambda \qquad \ldots(6)$$

[ '1' and '·' Represent differentiations w.r.t 'r' and t respectively ]

From equations (5) and (6) one can write

$$(C^1 / C) = (A^{11} / A^1) + A^1 \qquad \ldots(7)$$

Equation (7) readily integrates to yield

$$C = (e^A)^1 \qquad \ldots(8)$$

where the constant of integration has been absorbed by rescaling the radial coordinate without any loss of generality.

Now from eq.(5), by using eq.(8), we get

$$p[\ddot{b} + \tfrac{1}{2}(p+1)(\dot{b})^2] = e^{2A}[\tfrac{1}{2}p(p+1)(A^1)^2 + (p+1)\{(A^1)^2 + A^{11}\} - \Lambda] = b_0 \qquad \ldots(9)$$

where $b_0$ is the separation constant.

Now one can separate time and space parts as

$$[\ddot{b} + \tfrac{1}{2}(p+1)(\dot{b})^2] = (b_0/p) \qquad \ldots(10)$$

$$A^{11} + a(A^1)^2 = c + d\, e^{-2A} \qquad \ldots(11)$$

where $a = \tfrac{1}{2}(p+2)$, $c = [\Lambda/(p+1)]$ and $d = [b_0/(p+1)]$

In what follows we shall try to solve the system of equations for $b_0 = 0$ and for $b_0 \neq 0$.

## Case – I : $b_0 = 0$

In this case from eq.(10), we can solve b which becomes

$$e^b = t^{[2/(p+1)]} \qquad \ldots(12)$$

where the constant of integration has been absorbed by rescaling the time coordinate without any loss of generality.

Now one can integrate eq.(11) to yield

$$A = (1/a) \ln[\cosh(aH)r] \qquad \ldots(13)$$

where $H^2 = (c/a)$

using eq.(13), one gets from eq.(8) as

$$C = H[\sinh(aH)r][\cosh(aH)r]^{[(1-a)/a]} \qquad \ldots(14)$$

The string energy energy density F(r,t) can be found from eq.(4) which becomes

$$8\pi G\, F(r,t) = [\cosh(aH)r]^{(-2/a)}\, [\,\{\Lambda + 3aH^2 - 2a^2H^2 + 2apH^2\}$$

$$- \{2p(p-1)/(p+1)t^2\}\{\cosh(aH)r\}^{(-2/a)}$$

$$+ \{\tanh(aH)r\}^2\, \{H^2 - 3aH^2 + 2a^2H^2 - 2apH^2 + (3/2)pH^2 + \tfrac{1}{2}p^2H^2\}\,] \quad ..(15)$$

Finally the line element becomes

$$ds^2 = [\cosh(aH)r]^{(2/a)}\, [\,dt^2 - t^{[4/(p+1)]}\, dz_i^{\,2}\,] - dr^2$$

$$- H^2\,[\sinh(aH)r]^2\,[\cosh(aH)r]^{[2(1-a)/a]}\, d\theta^2 \quad\quad\ldots\ldots(16)$$

One should note that for $r \to 0$ i.e. near the axis of the string, the line element becomes

$$ds^2 = dt^2 - t^{[4/(p+1)]}\, dz_i^{\,2} - dr^2 - a^2 H^4 r^2\, d\theta^2 \quad\quad\ldots\ldots(17)$$

The non-static metric (17) shows that the proper volume becomes zero at $t \to 0$ and hence there is a disc like singularity in space-time.
Now we calculate curvature scalar, which gives

$$R = [\tanh(aH)r]^2\,[\,4aH^2 - 2a^2H^2 + 2apH^2 - (5/2)pH^2 - (3/2)H^2 - \tfrac{1}{2}p^2H^2\,]$$

$$- 4\,[\,2aH^2 - a^2H^2 + apH^2\,] \quad\quad\ldots..(18)$$

which is evidently time independent.
This result is similar to the case for non-static global string both in general relativity and in dilaton gravity and local string in Brans-Dicke theory in 4D cases [8].
It can be seen from (18) that the space-time becomes singular at different finite distances from the axis of the string for different p's i.e. p = 1, 2, 3, …….. etc.

# Case – II : $b_0 \neq 0$

In this case, we solve equations (10) & (11) to yield

$$e^b = [\cosh\{\tfrac{1}{2} E(p+1)\, t\}]^{[2/(p+1)]} \quad\quad\ldots..(19)$$

where $E^2 = [\,2 b_0 /\{p(p+1)\}\,]$

$$\int [\, D\, e^{-2A} + (c/a) + \{d/(a-1)\} e^{-2aA}\,]^{-\tfrac{1}{2}}\, dA = \pm(r - r_0) \quad\quad\ldots….(20)$$

where D and $r_0$ are integration constants.

From eq.(20), one can get solution of A in closed form only for D = 0.
Hence we get,

$$e^{2A} = L^2 \sinh^2(Hr) \qquad ....(21)$$

where $L^2 = [ad/\{c(a-1)\}]$, $H^2 = (c/a)$ and $r_0$ is taken as zero without any loss of generality.

The expression for C is

$$C = LH \cosh(Hr) \qquad ....(22)$$

The string density has the following form

$$8\pi G\, F(r,t) = L^{-2} [\operatorname{cosech}(Hr)]^2 [\Lambda + H^2 + 2pH^2 + \tfrac{1}{2} p(p-1)H^2 \coth^2(Hr)$$

$$- \tfrac{1}{2} p(p-1)E^2 \tanh^2 \tfrac{1}{2}\{E(p+1)t\}.\{\Lambda p/ b_0 (p+1)\} \operatorname{cosech}^2(Hr)] \qquad .....(23)$$

In this case the element is

$$ds^2 = L^2 \sinh^2(Hr) [dt^2 - \{\cosh \tfrac{1}{2} E(p+1)t\}^{[4/(p+1)]} dz_i^2] - dr^2$$

$$- H^2 L^2 \cosh^2(Hr) d\theta^2 \qquad ......(24)$$

The non-static metric (24) shows that the proper volume never vanishes for any value of t.
Thus the space-time is non-singular w.r.t. time.
This fact is reflected in curvature scalar given by

$$R = [\coth(Hr)]^2 [\{4p\Lambda/(p+2)\} + 2pH^2 - H^2(p^2 + 3p - 2)]$$

$$- [\{4p\Lambda/(p+2)\} + 2pH^2 + 2(p+2)H^2] \qquad .......(25)$$

which is evidently time independent.
Thus the proper volume and curvature scalar never vanish with time.
But from eq.(25), we see that the space-time becomes singular at finite distances from the axis of the string at different points for different 'p'.

In conclusion, we have found a class of exact interior solutions describing local cosmic string in a higher dimensional space. It is surprising to note that in case-I, the angular part becomes smaller if the number of coordinates 'p' of lower dimensional space-time increasing indefinitely. Whereas in case-II, if p is large enough, all the metric coefficients shrink with 'p'. Thus if one assumes that in the early stages of the universe, the dimension is much greater than 4, then one can see, in our model, either only angular part or the entire space-time had the size of planck's length $L_{pl} \sim 10^{-33}$ cm. Unlike the Arkani-Hamed et al [2] set up, the extra dimension here, are not at the millimeter scale but are infinite. In future, one should give a serious attention to the fact that local strings are alternative to domain walls in compactification processes.

# Acknowledgement:

F.R. is thankful to IUCAA for providing the research facilities.

# References:

[1]   Kaluza T   Sitz.Preuss.Akad.Wiss.Berlin.Math.Phys.K 1,966(1921);
      O.Klein  Z.Phys. 37,895(1926)
      S.Weinberg  Physics in higher dimension ( World Scientific , Singapore, 1986)

[2]   N. Arkani-Hamed et al Phys.Lett.B 429, 263 (1998); Phys.Rev.D 59, 086004
      (1999) ; R. Sundrum Phys.Rev.D 59, 085010 (1999)

[3]   L. Randall and R. Sundrum Phys.Rev.Lett. 83, 3370 (1999);
      Phys.Rev.Lett. 83, 4690 (1999)

[4]   A. Kohen and D. Kaplan Phys. Lett. B 470, 52 (1999)

[5]   R Gregory Phys.Rev.Lett. 84, 2564 (2000)

[6]   I. Olasagasti and A. Vilenkin Phys. Rev.D 62, 044014 (2000)

[7]   A. Vilenkin Phys. Rep. 121, 263 (1985)

[8]   R. Gregory Phys. Rev. D 54, 4955 (1996);
      O. Dando and R. Gregory , gr-qc / 9802015;
      A.A. Sen, gr-qc / 9803030